\documentstyle[pra,aps,epsf]{revtex}
\begin{document}

\title{
Large scale instabilities in two-dimensional magnetohydrodynamics
}

\author{G. Boffetta$^1$, A. Celani$^1$ and R. Prandi$^2$}
\address{$^1$ Dip. Fisica Generale, Universit\`a di Torino,
and INFM, Unit\`a di Torino Universit\`a , Italy} 
\address{$^2$ Dip. Fisica Teorica, Universit\`a di Torino,
and INFM, Unit\`a di Pisa, Italy} 

\date{\today}

\maketitle

\begin{abstract}
The stability of a sheared magnetic field
is analyzed in two-dimensional 
magnetohydrodynamics with resistive and viscous dissipation.
Using a multiple-scale analysis, it is shown that
at large enough Reynolds numbers the basic state describing a motionless
fluid and a layered magnetic field, becomes unstable
with respect to large scale perturbations.
The exact expressions for eddy-viscosity and eddy-resistivity are derived
in the nearby of the critical point where 
the instability sets in.
In this marginally unstable case the nonlinear phase 
of perturbation growth obeys to a Cahn-Hilliard-like dynamics
characterized by coalescence of magnetic islands leading to a final 
new equilibrium state.
High resolution numerical simulations confirm quantitatively the
predictions of multiscale analysis.
\end{abstract}


\section{Introduction}
\label{sec1}
Magnetic reconnection in two-dimensional
magnetohydrodynamics (MHD) 
is one of the most intriguing problems in plasma physics,
originally motivated by the observation of sudden and rapid release of energy
occurring for instance in solar flares and tokamak disruptions. These phenomena
are characterized by a change in the topology 
of the magnetic field lines and the
formation of island-like structures. Resistive instabilities 
are often addressed as responsible of magnetic reconnection
\cite{FKR63,PMW84}. 
Their development is a consequence of a resistive boundary layer formed nearby
a magnetic field neutral line, where the ideal constraint of frozen magnetic 
flux is relaxed. Their growth rates typically depend on 
fractional powers of the dissipative coefficients (through the dimensionless
inverse Lundquist number and Reynolds number). When nonlinear
effects are taken into account, the growth of resistive instabilities slows
down, eventually reaching saturation \cite{Bis93}. Numerical studies
\cite{Mal} have shown that the nonlinear phase of evolution is characterized by
the generation of large scale magnetic structures, by means of a coalescence
process of small magnetic flux patches.        
  
In this Paper we present a new kind of instability which sets in when the
influence of dissipative terms cannot be restricted to a limited domain
and the boundary layer approximation does not hold anymore.
This instability can be viewed as the MHD counterpart of the large scale 
hydrodynamical instability 
which is known to develop in highly
anisotropic flows \cite{Sin61}. 
Large scale instabilities are formally associated to the existence
of negative values of the eddy-viscosity and/or eddy-resistivity.
The exact values of the turbulent dissipative coefficients can be 
explicitly derived from the knowledge of the equations of motion and
of the basic equilibrium state, by means of multiple scale
analysis. In the case of a sheared magnetic field embedded in a motionless
fluid, a negative eddy-resistivity instability can develop for high enough 
Lundquist/Reynolds
numbers, causing the formation of a chain of magnetic islands. As observed for
resistive instabilities, also in this case, during the
nonlinear evolution, the number of magnetic islands decreases, due to the
merging of pairs of close islands. This coalescence process eventually leads to
the formation of a single island, whose width is determined by the largest scale allowed
in the system.

In Section \ref{sec2}, we briefly introduce the fluid equations of 
magnetohydrodynamics and their basic equilibrium states. In Section \ref{sec3},
the behavior of large scale perturbations is investigated making use
of multiple-scale analysis and exact expressions for eddy-viscosity
and eddy-resistivity are derived. The main result is the appearance 
of large scale transverse instabilities, associated to negative values 
of the renormalized dissipative coefficients. 
In Section \ref{sec4}, we focus on the 
case of marginal instability for which we obtain 
the effective equation 
for the large scale behavior and we show that the full nonlinear regime is
characterized by the evolution towards a fixed point.
In Section \ref{sec5}, we present  
the results of direct numerical simulations (DNS), which display a clear
{\it quantitative} agreement with the predictions of multiple-scale
analysis, both in the linear and in the nonlinear phase. 

\section{MHD equations and basic equilibria}
\label{sec2}
MHD equations are relevant in many different physical contests, such as
astrophysics, laboratory plasma physics or magnetized liquid metal dynamics. 
The applicability of this model, which is a fluid description of plasma,
relies on the assumption that
all the lengthscales under consideration must largely exceed 
the ion Larmor radius. 
In a strong external magnetic field which is oriented
along $z$, $B_z \gg B_{\perp}$, the motion becomes almost 
two-dimensional and the MHD model
is well approximated by the 2D MHD equations for the magnetic flux
function $\psi$ associated to the planar magnetic field 
(${\bf B}_{\perp}={\bf e}_z \times \nabla \psi$) and for the stream function 
$\varphi$ of the
incompressible planar flow (${\bf v}_{\perp}={\bf e}_z \times \nabla \varphi$)
\cite{Bis93}:
 
\begin{eqnarray}
\label{eq:psi}
\displaystyle{\partial\psi \over \partial t} + 
\left[\varphi,\psi\right] & = & \eta
(\nabla^2 \psi - J_0) \;,\\
\frac{\partial \nabla^2 \varphi}{\partial t} + 
\left[\varphi,\nabla^2 \varphi\right]
& = & \left[\psi,\nabla^2\psi\right] +  \nu \nabla^4 \varphi \;, 
\label{eq:phi}
\end{eqnarray}

where, following a standard notation, 
the convective terms are written as Jacobian
operators($\left[ f, g\right] = \partial_x f\partial_y g - \partial_x g
\partial_y f$).
The forcing $J_0=\nabla^2\psi_0$ represents an input of
magnetic energy preventing the decay due 
to resistive ($\eta$) and viscous ($\nu$) dissipation. 

The above equations have been normalized with respect to 
the characteristic macroscopic length $L$, magnetic field ${\bar B}$ and 
Alfv\`en time $\tau_A= L/v_A $, while $\eta$ and $\nu$ are, respectively, the
inverse Lundquist number ($\eta c^2/4\pi v_A L$) and the inverse Reynolds number
($\nu/v_A L$). The Alfv\`en
velocity, $v_A = {\bar B}/(4\pi m n)^{1/2}$, is the velocity of
small amplitude waves, propagating along the magnetic field ${\bar B}$ in a
uniform plasma with density $n m$. 

The equations (\ref{eq:psi},\ref{eq:phi}) admit a  basic equilibrium 
state $\psi=\psi_0={\cal F}(x),\varphi=0$, where ${\cal F}$ is any function.
We consider ${\cal F}(x)= \cos x$, in order to investigate the stability of a 
sheared magnetic field in a motionless conducting fluid, in a slab geometry with 
periodic boundary conditions. Such a configuration is widely used 
to study the evolution of reconnecting modes.
It is known \cite{Bis93} that the slab geometry can be trusted as
realistic and resistive instabilities can develop only if the slab 
aspect ratio ($L_x/L_y$) is less than one. For this reason it is of 
great interest the question related to the stability of the above 
configuration with respect to 
transverse perturbations on scales larger than the 
typical scale of the basic magnetic flux. 
These large scale instabilities are eventually responsible for the onset 
of the inverse cascade of the square magnetic potential $A=\langle
\psi^2\rangle$ in fully developed 2D
MHD turbulence \cite{Bisk}.

\section{Multiple-scale analysis}
\label{sec3}
As a first step, using a standard multiple-scale technique, 
we derive analytically the large-scale magnetic flux $\Psi$ and stream 
function $\Phi$ 
equations. The multiscale method is based on the idea of exploiting the
separation of scales as a perturbative parameter.  
We consider a periodic box width $L_{box}$ much 
larger than the basic magnetic flux typical length-scale ($L_{box} \gg L_x$) in
order to follow the dynamics on spatial scales of order 
$O(1/\varepsilon)L_x$. Beside the {\it fast} variables ($x,y,t$) on which the 
basic flow evolves, a set of {\it slow} variables ($X=\varepsilon x,
Y=\varepsilon y, T=\varepsilon^2 t$) can be introduced.
According to this choice the differential operators appearing in 
(\ref{eq:psi},\ref{eq:phi}) are transformed to
\begin{equation}
\partial_i \rightarrow \partial_i + \varepsilon \nabla_i \;, \hspace{20pt}
\partial_t \rightarrow \partial_t + \varepsilon^2 \partial_T \;.
\label{eq:der}
\end{equation}
Expanding perturbatively in $\varepsilon$ the fields we obtain:
\begin{eqnarray}
\nonumber
\psi&=&\psi^{(0)}(x,X,y,Y,t,T)+\varepsilon \psi^{(1)}(x,X,y,Y,t,T)+\varepsilon^2
\psi^{(2)} ... \\
\varphi&=&\varphi^{(0)}(x,X,y,Y,t,T)+\varepsilon \varphi^{(1)}(x,X,y,Y,t,T)+
\varepsilon^2\varphi^{(2)} ... 
\label{eq:exp}
\end{eqnarray}
The scaling of the {\it slow} time $T$ is suggested by physical hints: we are 
looking for a
diffusive behavior of large scales which takes place on times
$O(\varepsilon^{-2})$. 
It is worth noticing that in general the large-scale MHD dynamics is first 
order in time and space (the well-known $\alpha$ effect) \cite{Mof78}, but it 
can be shown that this is not the case for parity-invariant basic 
configurations \cite{Dub91}.

By substituting (\ref{eq:der}) and (\ref{eq:exp}) 
in (\ref{eq:psi}) and (\ref{eq:phi}) and by equating 
the same power of $\varepsilon$, one easily finds
a hierarchy of equations in which perturbations belonging to different order of
expansion appear coupled and depend on {\it fast} and {\it slow} variables.
The dependence on the fast time variables can be discarded 
 by observing that it reduces to a
transient not affecting the long-time behavior, 
 thanks to the fact that the forcing
and the basic flux are time-independent 
(a rigorous proof needs the
construction of a Poincar\`e inequality).
We look for solutions with the same periodicities of the 
basic magnetic flux in $L_{box}$. At each step we can distinguish the pure
large field
contribution from the small scale oscillating part:
\begin{displaymath}
\psi^{(k)}=\Psi^{(k)}(X,Y,T)+\large{\tilde{\psi}}^{(k)}(X,Y,T,x) 
\; , \hspace{10pt}
\varphi^{(k)}=\Phi^{(k)}(X,Y,T)+\displaystyle{\tilde{\varphi}}^{(k)}(X,Y,T,x)
\;.
\end{displaymath}
Equations have to be solved recursively because solutions of lower order appear
as coefficients in the following steps of the hierarchy. At each order one has
to test the validity of the solvability condition. Indeed, the 
equation for the large scale magnetic flux $\Psi^{(0)}$ is obtained as
solvability condition at order $\varepsilon^2$,
while the equation for the large scale vorticity $\nabla^2\Phi^{(0)}$ 
comes out at order $\varepsilon^4$:

\begin{eqnarray}
\label{eq:PSI}
\frac{\partial \Psi^{(0)}}{\partial T} +
\displaystyle{\left[\Phi^{(0)},\Psi^{(0)}\right]}& = & \eta
\nabla^2 \Psi^{(0)} - \frac{1}{2 \nu}\frac{\partial^2 \Psi^{(0)}}{\partial Y^2}
\;, \\
\nonumber
\frac{\partial \nabla^2 \Phi^{(0)}}{\partial T} + \left[\Phi^{(0)},\nabla^2 
\Phi^{(0)}\right]&=&
\left[\Psi^{(0)},\nabla^2\Psi^{(0)}\right] +
\frac{1}{2}\frac{\partial^2}{\partial X\partial Y}
\left\{
\frac{1}{\nu^2}\left(1+2\frac{\nu}{\eta}\right)\left(\frac{\partial\Psi^{(0)}}
{\partial Y}\right)^2+
\frac{1}{\eta^2}\left(\frac{\partial \Phi^{(0)}}{\partial Y}\right)^2\right\}
\\
\label{eq:PHI}
&&+\frac{1}{2\eta}\frac{\partial^2}{\partial Y^2}\left(\frac{\partial^2}
{\partial Y^2}-3\frac{\partial^2}{\partial X^2}\right)\Phi^{(0)} + 
\nu \nabla^4 \Phi^{(0)} \; .
\end{eqnarray}

Let us focus our attention on diffusive terms in (\ref{eq:PSI}) and
(\ref{eq:PHI}): as a consequence of the anisotropy of the basic small-scale
flow the eddy-diffusivities are anisotropic too.
For longitudinal perturbations 
($\Phi^{(0)}=\Phi^{(0)}(X,T)$,$\Psi^{(0)}=\Psi^{(0)}(X,T)$)
both viscosity 
and resistivity are left unchanged. On the other hand, for transverse
perturbations  
($\Phi^{(0)}=\Phi^{(0)}(Y,T)$,$\Psi^{(0)}=\Psi^{(0)}(Y,T)$)
the renormalization of resistivity due to the small scale
magnetic energy holds a negative term ($-1/2\nu$),
while molecular viscosity is increased by the eddy contribution ($1/2\eta$).
 
It is interesting to notice that the possibility of a negative eddy-resistivity
due to small scale magnetic energy has already been presented in \cite{Pou78}.
That result, obtained in the framework of closure approaches to 2D MHD
turbulence, was suggested as an explanation of the square magnetic potential
inverse cascade.
Analogous results are reported in
\cite{Bis84}, where the effective resistivity is shown to become negative in a
small scale turbulent plasma, as long as the magnetic energy exceeds the kinetic
one.    

By inspection of equations (\ref{eq:PSI}) and
(\ref{eq:PHI}), we stress that the large scale magnetic flux and stream
function are linearly decoupled. If we assume large scale 
perturbations of the type 
$\Psi^{(0)} \sim \exp(\Gamma_{\Psi} T + \imath K_X X +\imath K_Y Y)$ and 
$\Phi^{(0)} \sim \exp(\Gamma_{\Phi} T + \imath K_X X + \imath K_Y Y)$, stability
analysis leads to the following dispersion relationships

\begin{eqnarray}
\label{gammapsi}
\Gamma_{\Psi}&=& -\eta K_X^{2} - (\eta- \frac{1}{2\nu}) K_Y^{2} \;,\\
\Gamma_{\Phi}&=&-\frac{1}{K^2}\left[(\nu+\frac{1}{2\eta})K_Y^4
+ 2 ( \nu-\frac{3}{4\eta})K_X^2K_Y^2 +\nu K_X^4\right] \;.
\label{gammafi}
\end{eqnarray}

\begin{figure}[ht]
\epsfxsize=220pt\epsfysize=200pt\epsfbox{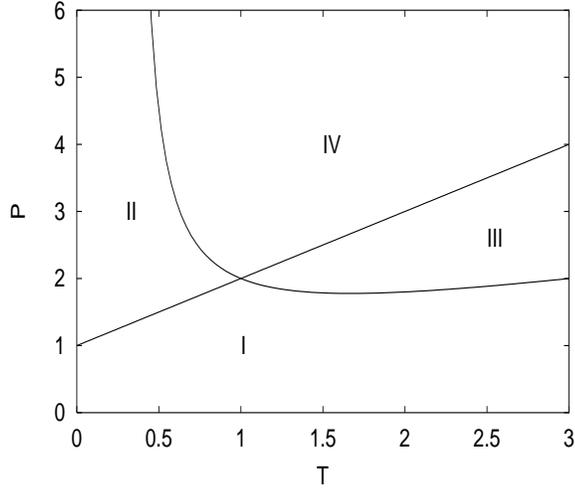}
\caption{
Marginal stability lines for magnetic flux and stream function.
Region $I$: stable with $\Gamma_{\Psi} <0$ and $\Gamma_{\Phi} <0$.
Region $II$: $\Gamma_{\Psi} > 0$, $\Gamma_{\Phi} < 0$.
Region $III$: $\Gamma_{\Psi} < 0$, $\Gamma_{\Phi} > 0$.
Region $IV$: instable with $\Gamma_{\Psi} > 0$, $\Gamma_{\Phi} > 0$.
}
\label{fig0}
\end{figure}

The stability problem can be tackled more easily by introducing the parameters
$P=1/2\eta\nu$ and $T=(K_X/K_Y)^2$. Marginal stability lines
($\Gamma_{\Psi}=\Gamma_{\Phi}=0$) are then given by 

\begin{eqnarray*}
&& 1-P + T = 0 \;, \\
&& 1+P + 2 (1- \frac{3}{2}P)\;T + T^2 = 0 \;.
\end{eqnarray*}

and plotted in Figure \ref{fig0}.

For high enough values of molecular resistivity and viscosity 
($P < 1$), the basic flow is stable against
any large scale perturbation. Increasing the Reynolds numbers, the first
instability sets in at $P=1$ and $T=0$, i.e. for transverse perturbations.  
We notice that for $1 < P < 16/9$ the large scale vorticity is always stable
($\Gamma_{\Phi} < 0$) and the magnetic potential growth rate $\Gamma_{\Psi}$ is
maximum in the case $T = 0$.

In the following, we investigate the linear and nonlinear evolution of large
scale perturbations in the neighborhood of the critical point
$(P=1, T=0)$, where we expect that
there is no amplification of kinetic energy but that due to
non-linear effects of coupling between velocity and magnetic field.
 
\section{Marginal instability}
\label{sec4}
Special attention deserves the development of this large scale instability 
for Reynolds numbers close to the marginal stability threshold.
In this regime, it is possible indeed to follow the full nonlinear evolution of
the perturbation. We show that the instability eventually reaches a fixed point
characterized by a magnetic island of the size of the box. The nonlinear
evolution in the marginal regime is described by two coupled equations which
generalize the Cahn-Hilliard equation, found for the hydrodynamical counterpart
of this system, the so-called Kolmogorov flow \cite{Dub91,Siv85,She87}. 

Let us suppose to move the parameters just above the marginal stability 
line: 
\begin{equation}
\eta=\eta_c(1-\varepsilon^2) \;, \hspace{10pt} \nu=\nu_c(1-\varepsilon^2)
\label{eq:par}
\end{equation}
where $\eta_c\nu_c=1/2$ ($P=1$). 
The perturbative parameter $\varepsilon$ is
thus fixed by the distance between $\eta,\nu$ and their critical values
$\eta_c, \nu_c$. We will take into account only transverse perturbations,
since, as shown in Figure \ref{fig0}, the large scale magnetic flux linear 
instability is mainly transverse, close to $P=1$. According to 
(\ref{eq:PSI}), the transverse eddy-resistivity
in the neighborhood of the critical line defined by (\ref{eq:par})
is of order $O(\varepsilon^2)$, thus suggesting a scaling
for the slow time $T=\varepsilon^4 t$.
The decomposition rules (\ref{eq:der}) become  
\begin{equation}
\partial_x \rightarrow \partial_x \; \hspace{5pt}
\partial_y \rightarrow \partial_y+\varepsilon\partial_Y \;, \hspace{5pt}
\partial_t \rightarrow \partial_t + \varepsilon^4\partial_T \; .
\label{eq:der2}
\end{equation}

The same multiscale technique described above can be adopted to solve  
perturbatively (\ref{eq:psi}) and (\ref{eq:phi}). 

At first order in $\varepsilon$ we obtain:
\begin{eqnarray}
\label{eq:psi1}
\psi&=&\cos x + \Psi^{(0)}(Y,T)+\varepsilon\Psi^{(1)}(Y,T) \\ 
\varphi&=& 2 \eta_c \varepsilon \frac{\partial \Psi^{(0)}}
{\partial Y} \sin x \; .
\label{eq:phi1}
\end{eqnarray}
We notice that, according to the conclusions drawn above, the large scale
stream function is linearly stable and it is simply driven by 
the magnetic flux. 

The evolution equations for $\Psi^{(0)}$ and $\Psi^{(1)}$ emerge as solvability
conditions at order $\varepsilon^4$ and $\varepsilon^5$:

\begin{eqnarray}
\label{eq:psi0p}
\partial_T
\Psi^{(0)}&=&-\frac{27}{8}\eta_c\partial_{4Y}\Psi^{(0)}-2\eta_c\partial_{YY}
\Psi^{(0)}+12\eta_c\partial_{YY}\Psi^{(0)}(\partial_Y \Psi^{(0)})^2 \\
\label{eq:psi1p}
\partial_T
\Psi^{(1)}&=&-\frac{27}{8}\eta_c\partial_{4Y}\Psi^{(1)}-2\eta_c\partial_{YY}
\Psi^{(1)}+12\eta_c\partial_{YY}\Psi^{(1)}(\partial_Y \Psi^{(0)})^2 \\
\nonumber
&&+24\eta_c\partial_Y\Psi^{(0)}\partial_{YY}\Psi^{(0)}\partial_Y\Psi^{(1)}
\end{eqnarray}

In the first equation (\ref{eq:psi0p}), one easily recognizes 
the renowned Cahn-Hilliard equation \cite{Siv85},
which may be written in variational form:
\begin{displaymath}
\frac{\partial \Psi^{(0)}}{\partial t}= -\frac{\delta V\left[\Psi^{(0)}\right]}
{\delta \Psi^{(0)}} \; .
\end{displaymath}
The existence of the Lyapunov functional 
\begin{equation}
\label{eq:lya}
V\left[\Psi^{(0)}\right]=
\eta_c\int dY \left[-(\partial_Y\Psi^{(0)})^2 + (\partial_Y \Psi^{(0)})^4
+\frac{27}{16}(\partial_{YY}\Psi^{(0)})^2\right]
\end{equation} 
indicates that asymptotically the solution of (\ref{eq:psi0p}) in a bounded
domain reaches a fixed point.  
This stationary solution is approached by a sequence of
metastable states of decreasing dominating mode. We thus expect to observe a
nonlinear evolution dominated by a magnetic island coalescence, analogous to
the vortex pairing in 2D hydrodynamics \cite{She87}.

Equation (\ref{eq:psi1p}) is linear in $\Psi^{(1)}$, with coefficients 
depending nonlinearly on $\Psi^{(0)}$.
It also can be written as a gradient flow, with a Lyapunov
functional
\begin{equation}
\label{eq:lya2}
V\left[\Psi^{(1)}\right]=
\eta_c\int dY \left[-(\partial_Y\Psi^{(1)})^2 + 
6 (\partial_Y \Psi^{(0)})^2 (\partial_Y \Psi^{(1)})^2
+\frac{27}{16}(\partial_{YY}\Psi^{(1)})^2\right] \;.
\end{equation} 

We conclude this section by observing that the dispersion relation for
$\Psi^{(0)} \sim \exp(\Gamma T + \imath K Y)$ now reads:
\begin{equation}
\label{eq:reldisp}
\Gamma=-\frac{27}{8}\eta_cK^4+2\eta_cK^2 \;.
\end{equation}
It implies instability ($\Gamma < 0$) for any ($K \stackrel{\sim}{<} 0.77$).
We notice that information about the characteristic scale of unstable modes
were totally absent in the general treatment presented in the previous section
(see equation (\ref{gammapsi}).

\section{Numerical results}
\label{sec5}
The results obtained in  the previous section have been checked 
by extensive direct numerical 
simulations of MHD equations (\ref{eq:psi},\ref{eq:phi}). 
In order to force a transverse perturbation, we integrate the equations
on a rectangular slab with $L_{box}=L_{x}=2 \pi$ and $L_{y} \ge L_{x}$.
In this way large scale instability can only develop on the $y$
direction for an aspect ratio $r=L_x/L_y<1$.

Given the numerical values of parameters $\eta$ and $\nu$, from 
(\ref{eq:par}) and the condition $\eta_c \nu_c = 1/2$ we have 
\begin{equation}
\varepsilon = \sqrt{1-\sqrt{2 \eta \nu}} \;, \hspace{10pt}
\eta_c = \sqrt{{\eta \over 2 \nu}}
\label{eq:parsim}
\end{equation}
which are used for the theoretical predictions of the previous section.

The simplest check of our predictions concerns the growth rates of the
instability which, in the initial linear regime, are given by the
dispersion relation (\ref{eq:reldisp}). In physical (not rescaled)
variables (\ref{eq:reldisp}) becomes
\begin{equation}
\gamma=-\frac{27}{8}\eta_c k^4+2\eta_c \varepsilon^2 k^2
\label{eq:reldisp2}
\end{equation}
which shows that the largest unstable wavenumber is 
$k_{max} \simeq 0.77 \varepsilon$. 
The smallest transverse wavenumber is $k_{1}=r$ thus, in order
to numerically observe the instability, it must be
$r \le 0.77 \varepsilon$.

\begin{figure}[ht]
\epsfxsize=220pt\epsfysize=200pt\epsfbox{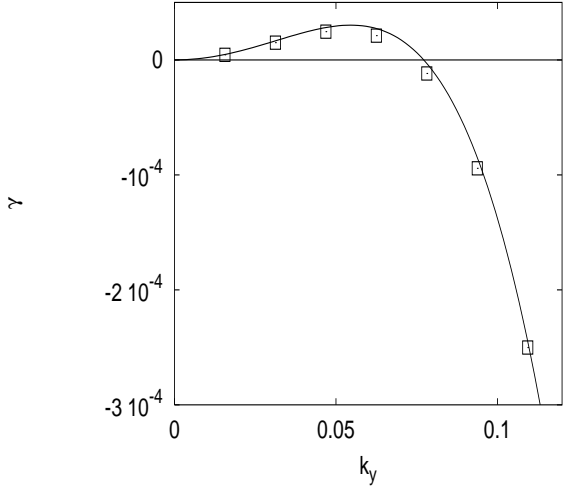}
\caption{
Growth rates $\gamma$ of the transverse Fourier modes $k$ for 
simulation with $r=1/64$, $\nu=0.49$ and $\eta=1.0$. The 
continuous line represent the linear prediction (\ref{eq:reldisp2}).
}
\label{fig1}
\end{figure}

In Figure \ref{fig1} we report the growth rates of the first 
modes for a simulation with $r=1/64$, $\nu=0.49$ and $\eta=1$.
We have $\varepsilon \simeq 0.1$ and thus only the first $4$ modes
are unstable. The initial perturbation is small, random and on
all the first $20$ modes, thus we are able to observe also negative
$\gamma$'s (stable modes). The comparison with the linear prediction
(\ref{eq:reldisp2}) is very good even for $\varepsilon$ not very small.
The numerical data of Figure \ref{fig1} are obtained by a linear fit
of the logarithm of the mode amplitude versus time
in the early stages of the simulation. 

\begin{figure}[ht]
\epsfxsize=220pt\epsfysize=200pt\epsfbox{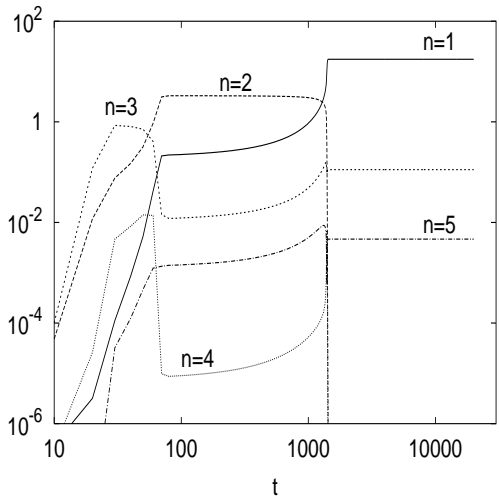}
\caption{
Time evolution of magnetic potential of the first Fourier transverse
components of wavenumber $k_n=n/16$ for the DNS with $r=1/16$, 
$\eta=0.4$ and $\nu=1.0$.
The number of unstable modes is $4$. 
}
\label{fig2}
\end{figure}

Let us now consider the nonlinear stage of the perturbation growth.
We describe here a different simulation with $r=1/16$ and 
$\varepsilon \simeq 0.32$ which was advanced for a very long lapse of time.
The nonlinear evolution will ultimately lead to a fixed point 
by a succession of long lasting quasi equilibrium states of decreasing
wavenumber. The evolution of the amplitudes of the first $5$ transverse 
modes computed from the direct numerical simulation is plotted in
Figure \ref{fig2}. Observe that in this case the fifth mode
$k=5/16$ is linearly stable but it is non-linearly driven by smaller
wavenumber. 

\begin{figure}[ht]
\epsfxsize=220pt\epsfysize=200pt\epsfbox{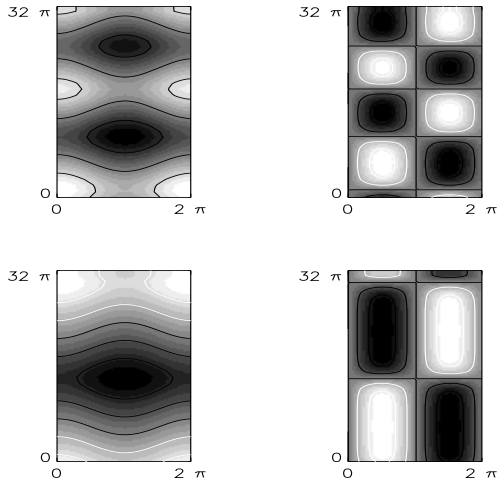}
\caption{
Snapshot of the magnetic flux $\psi$ (left) and stream function
$\varphi$ (right) for $t=200$ (upper) and $t=20000$ (lower).
}
\label{fig3}
\end{figure}

The typical linear time is now rather short, $1/\gamma \sim O(1)$, and
the final stationary state, dominated by the largest available mode
$k_1$, is reached at very long times, $t > 1000$. At intermediate times,
almost stationary metastable states, characterized by decreasing
leading mode, are punctuated by fast coalescence processes.
Most of the energy dissipation takes place during this fast
reconnection processes.
In Figure \ref{fig3} we display the period-two
metastable state at time $t=200$ and the final, period-one state
at $t=20000$. 
The dynamical picture arising from Figures \ref{fig2} and \ref{fig3}
qualitatively agrees with the dynamics described by the Cahn-Hillard 
equation \cite{She87}.

\begin{figure}[ht]
\epsfxsize=220pt\epsfysize=200pt\epsfbox{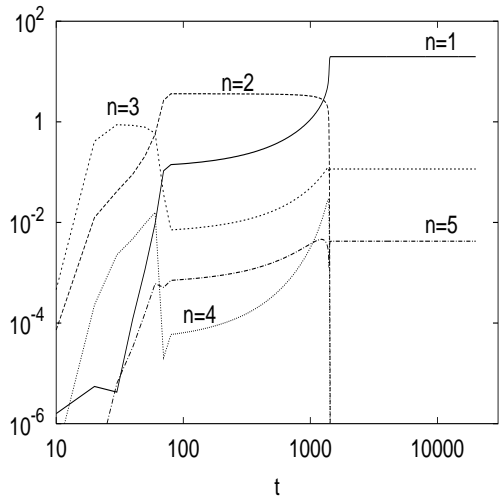}
\caption{
Time evolution of the first square Fourier components of $\Psi^{(0)}$
solution of the Cahn-Hillard equation. Compare with Figure \ref{fig2}
relative to the direct numerical simulation of MHD equations.
}
\label{fig4}
\end{figure}

To check quantitatively the validity of non linear multiscale analysis we have
numerically integrated the Cahn-Hilliard equation for the 
large scale magnetic flux (\ref{eq:psi0p}) with the same parameters
of the DNS. As shown in Figure \ref{fig4} we find an impressive
agreement even for very long times.
The final relative amplitudes of the most energetic transverse
modes is recovered within a $10 \, \%$ accuracy.

\begin{figure}[ht]
\epsfxsize=220pt\epsfysize=200pt\epsfbox{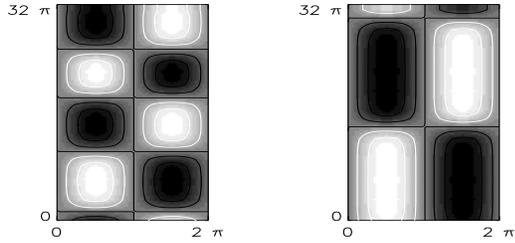}
\caption{
Snapshot of the stream function $\varphi$ reconstructed according to 
(\ref{eq:phi1}) for $t=200$ (left) and $t=20000$ (right).
}
\label{fig5}
\end{figure}

As a further test of the multiscale predictions, we checked the 
relations (\ref{eq:psi1}-\ref{eq:phi1}) during the evolution.
At leading order in $\varepsilon$, $\Psi^{(0)}(y,t)$ is obtained
by subtracting the basic flow $\cos x$ from the magnetic flux $\psi(x,y,t)$.
The resulting field, which reveals to be indeed $x$-independent,
is then used to reconstruct the stream function by means of (\ref{eq:phi1}).
The results for the configuration of Figure \ref{fig3} is shown in
Figure \ref{fig5}.

\section{Conclusions}
\label{sec6}
We have investigated the issue
of stability of highly anisotropic, magnetic-energy dominated 
equilibrium states of the MHD fluid model equations.
These configurations are known to be unstable for small values of resistivity
leading to the formation of thin boundary layers in the nearby of the 
neutral line of the magnetic field. At variance with
the above case, we focused our attention on 
the range of moderate Lundquist/Reynolds numbers, where the boundary layer 
approximation is not fruitfully applicable. 
In this situation we have shown analytically, 
by means of multiple-scale analysis, 
that large scale instabilities can arise above a definite threshold, and that
for a generic perturbation the maximum growth is achieved 
by modes transverse to the magnetic field lines of the basic state.
On the basis of this result, we have performed the multiple scale
analysis for the marginally unstable case and for purely transverse
perturbations. The analytic procedure yields
a couple of partial differential equations
which describe the full nonlinear evolution of magnetic perturbations. 
It is possible to show that these equations possess a Lyapunov functional
and thus their solutions asymptotically approach a fixed point
which represents a nonlinear equilibrium different from the basic one.
The kinetic perturbations are always linearly stable, and their growth is
uniquely due to the nonlinear coupling to the magnetic field.
Numerical simulations of two-dimensional MHD performed with a pseudospectral 
code reveal an excellent {\it quantitative agreement} with the
first-order analytical prediction in a wide range of values of 
the perturbative parameter. 

We therefore conclude that the loss of stability of parallel magnetic field
configurations at moderate Reynolds number is due to the
growth of large scale perturbations, and that the features of this 
instability can be captured by a multiple scale analysis.
The transverse large-scale instability is likely to be the
generic mechanism of instability of sheared magnetic fields
even for large Lundquist/Reynolds numbers whenever the basic state admits
a large number of neutral lines. 
When there is a single neutral line at large enough Lundquist/Reynolds 
numbers this mechanism is overcome by the formation of resistive 
boundary layers.

\section{Acknowledgments}
The authors wish to acknowledge support and hospitality
by the Istituto di Cosmogeofisica CNR, Torino.  
The calculations were partially performed with computer facilities
of INFN, Sezione di Torino.



\newpage

\end{document}